\documentclass[
    a4paper, 
    aps,
    pra,
    amssymb, 
    amsmath, 
    superscriptaddress, 
    amsfonts, 
    reprint, 
    showkeys, 
    nofootinbib, 
]{revtex4-2}

\usepackage[english]{babel}
\usepackage[utf8]{inputenc}
\usepackage[left=23mm,right=13mm,top=24mm,bottom=25mm,columnsep=15pt]{geometry} 
\usepackage[T1]{fontenc}
\usepackage{mathtools}
\usepackage{physics}
\usepackage{import}
\usepackage[table]{xcolor}
\usepackage{graphicx}
\usepackage{csquotes}
\usepackage{multirow}
\usepackage{rotating} 
\usepackage{booktabs}
\usepackage{adjustbox}
\usepackage{fontawesome}
\usepackage{dcolumn}
\usepackage{dsfont}
\usepackage{placeins}
\usepackage{epstopdf}
\usepackage[titletoc,toc,page,title,header]{appendix}
\usepackage[pdftex, pdftitle={Article}, pdfauthor={Author}]{hyperref}
\usepackage{orcidlink}
\usepackage{xspace}
\usepackage{xcolor}
\usepackage{multirow}
\usepackage{tcolorbox}
\usepackage{datetime}
\usepackage{tabularx}
\usepackage{makecell}
\usepackage{caption}
\usepackage{CJKutf8}
\usepackage{markdown}
\usepackage{listings}
\tcbuselibrary{listings}
\newdateformat{monthyeardate}{%
  \monthname[\THEMONTH] \THEYEAR
}

\definecolor{codebkg}{HTML}{ebebeb}

\hypersetup{
    colorlinks=true,
    linkcolor=blue,
    filecolor=magenta,      
    urlcolor=blue,
    citecolor=blue
}

\begin{document}

\title{Quantum resources in resource management systems}

% \author{Quantum \& HPC Working Group}
% Bacher
\author{Utz Bacher\,\orcidlink{0009-0007-6044-290X}}
\affiliation{IBM Deutschland Research \& Development GmbH, Boeblingen, Germany}

% Birmingham
\author{Mark Birmingham}
\affiliation{The Hartree Centre, STFC, Sci-Tech Daresbury, Warrington, WA4 4AD, United Kingdom}

% Carothers
\author{Christopher D. Carothers}
\affiliation{Center for Computational Innovation, Rensselaer Polytechnic Institute,  Troy, NY USA 12180}

% Damin
\author{Andrew Damin}
\affiliation{Center for Computational Innovation, Rensselaer Polytechnic Institute,  Troy, NY USA 12180}

% Gonzalez Calaza
\author{Carlos D. Gonzalez Calaza\, \orcidlink{0000-0001-8467-8633}}
\affiliation{Jülich Supercomputing Centre, Forschungszentrum Jülich, 52425 Jülich, Germany}
\affiliation{RWTH Aachen University, 52056 Aachen, Germany}

% Kumar Karnad
\author{Ashwin Kumar Karnad\, \orcidlink{0000-0001-8162-1957}}
\affiliation{Jülich Supercomputing Centre, Forschungszentrum Jülich, 52425 Jülich, Germany}

% Mensa
\author{Stefano Mensa\,\orcidlink{0000-0002-0938-144X}}
\affiliation{The Hartree Centre, STFC, Sci-Tech Daresbury, Warrington, WA4 4AD, United Kingdom}

% Moreau
\author{Matthieu Moreau\,\orcidlink{0009-0001-7886-3426}}
\affiliation{PASQAL, 24 Av. Emile Baudot, 91120 Palaiseau, France}

% Nober
\author{Aurelien Nober\,\orcidlink{0009-0006-8442-1401}}
\affiliation{PASQAL, 24 Av. Emile Baudot, 91120 Palaiseau, France}

% Ohtani
\author{Munetaka Ohtani\,\orcidlink{0009-0000-5286-0393}}
\affiliation{IBM Quantum, IBM Research – Tokyo, Tokyo 103-8510, Japan}

% Rossmannek
\author{Max Rossmannek\,\orcidlink{0000-0003-1725-9345}}
\affiliation{IBM Quantum, IBM Research—Zurich, Säumerstrasse 4, 8803 Rüschlikon, Switzerland}

% Rubin
\author{Philippa Rubin}
\affiliation{The Hartree Centre, STFC, Sci-Tech Daresbury, Warrington, WA4 4AD, United Kingdom}

% Emre
\author{M.~Emre~Sahin\,\orcidlink{0000-0002-5996-0407}}
\affiliation{The Hartree Centre, STFC, Sci-Tech Daresbury, Warrington, WA4 4AD, United Kingdom}

% Oscar
\author{Oscar Wallis\,\orcidlink{0009-0002-7323-2059}}
\affiliation{The Hartree Centre, STFC, Sci-Tech Daresbury, Warrington, WA4 4AD, United Kingdom}

% Shehata
\author{Amir Shehata}
\affiliation{Oak Ridge National Laboratory, Oak Ridge, TN 37831, USA}

% Sitdikov
\author{Iskandar Sitdikov\,\orcidlink{0000-0002-6809-8943}}
\affiliation{IBM Quantum, IBM T.J. Watson Research Center, Yorktown Heights, NY 10598, USA}

% Wennersteen
\author{Aleksander Wennersteen\,\orcidlink{0009-0006-5486-0980}}
\affiliation{PASQAL, 24 Av. Emile Baudot, 91120 Palaiseau, France}

%TBC
\date{\today}

% Contents
% Intro
% Background
% - resource management systems
% - quantum systems (hardware boxes)
% - vendor specific terms
% Design
% - quantum resource abstraction model
% - on-node vs shared resource model
% - direct link vs cloud bursting
% - QRMI design
% - full architecture
% Implementation
% - QRMI
% - spank plugin
% - gres
% Usage
% - example workflow (SQD iterative + optimal resource usage)
% - AiMOS usage
% - STFC chemistry cloud usage
% Appendix
% - resource management in container management systems

\begin{abstract}

Quantum computing resources are increasingly being incorporated into high-performance computing (HPC) environments as co-processors for hybrid workloads. To support this paradigm, quantum devices must be treated as schedulable first-class resources within existing HPC infrastructure. This enables consistent workload management, unified resource visibility, and support for hybrid quantum-classical job execution models.

This paper presents a reference architecture and implementation for the integration of quantum computing resources, both on-premises and cloud-hosted into HPC centers via standard workload managers. We introduce a Slurm plugin designed to abstract and control quantum backends, enabling seamless resource scheduling, minimizing queue duplication, and supporting job co-scheduling with classical compute nodes. The architecture supports heterogeneous quantum resources and can be extended to any workload (and container) management systems.

% Quantum computers are beginning to operate in high-performance computing (HPC) environments. Quantum can complement classical resources for specific workloads, but their adoption depends on integration into existing HPC infrastructure. Treating quantum devices as first-class resources allows for unified scheduling, improved usability, and support for hybrid quantum-classical applications.

% This paper presents the design architecture and reference implementation for quantum resources control using existing workload management systems. We introduce a suite of plugins for Slurm that enable integration of on-prem and cloud quantum computing resources into existing high-performance computing  centers. The paper details the interface design, plugin concept and implementation, operational aspects for heterogeneous compute clusters, as well as considerations for other resource management systems.

\end{abstract}

\maketitle

\section{Introduction}

The emergence of quantum computing motivates the incorporation of quantum resources into established high-performance computing (HPC) environments as a way of unlocking previously intractable workflows.
The integration of quantum computing into HPC environments is an active research area~\cite{ALEXEEV2024666, BECK2024161, QMIO2025, MANSFIELD2025}, where the development of appropriate software abstractions turns out to be a hurdle.
Hybrid quantum-classical applications, such as Sample-Based Quantum Diagonalization (SQD)~\cite{sqd}, are already being executed on large-scale HPC systems. Production HPC centers typically rely on a single workload management system to coordinate all resources, in order to avoid the operational overhead of multiple scheduling layers (including multiple queues), incosistent management, and degraded user workflows.

The differences between existing workload manager resource abstractions and the heterogeneous interfaces of quantum computers, whether locally hosted or accessed through cloud services, require an intermediate compatibility layer. Earlier approaches commonly used wrapper scripts, external submission mechanisms, or relied on specific software stacks~\cite{mqss}.
%which created operational inefficiencies (such as multi-layer scheduling) \textcolor{red}{[AW: IDk if I agree with this being thrown in without further expansion and sources]}.
A primary challenge in integrating quantum hardware is the diversity of platforms and their vendor-specific APIs. A solution tied to a single vendor's control system would be brittle and limit future flexibility.
The Quantum Resource Management Interface (QRMI), which we present in this work, is a conceptual framework designed to solve this problem by providing a standardized, vendor-agnostic abstraction layer, without being coupled to any single software stack. The design follows a loosely coupled architecture, allowing adoption across different resource managers and quantum providers, and support for both on-prem and cloud-based resources.

As a reference implementation, we target Slurm~\cite{slurm_overview}, a widely deployed workload manager in HPC. Slurm’s plugin framework provides the necessary hooks to extend its scheduling and resource management capabilities. We introduce a plugin that registers quantum resources as schedulable entities, enabling users to submit and manage hybrid jobs within the same Slurm-managed environment. The same architectural concepts can be generalized to other workload management systems.

Finally, we present integration architectures deployed in operational data centers that have adopted this work, demonstrating feasibility and evaluating system behavior in production environments.

%\textcolor{red}{The rest of this paper is structured as follows}

% ########
% old text
% ########

% The advent of quantum computing necessitates the integration of quantum resources into traditional HPC environments. Hybrid quantum–classical applications increasingly run on large HPC systems. Production sites prefer to manage compute resources via single workload management system and avoid problem of multiple scheduling systems. The gap between workload management compute resource abstractions and local or cloud-hosted quantum computers motivates a thin compatibility layer. Previous efforts relied on wrapper scripts, while our contribution is, to the best of our knowledge, the first to integrate at native to resource management systems, making quantum computers \emph{first-class} scheduler entities. 

% For reference implementation we choose Slurm. Slurm, a widely adopted workload manager, offers the SPANK framework to extend its capabilities. This work introduces SPANK plugins \footnote{\href{https://github.com/qiskit-community/spank-plugins/}{\textcolor[HTML]{840484}{\faGithub}~{https://github.com/qiskit-community/spank-plugins/}}} designed to manage quantum resources, enabling users to submit and control quantum jobs alongside classical tasks within Slurm-managed clusters. We also describe how to expand integration architecture to other resource management systems.

% The context diagram in figure \ref{fig:context} provides an overview of involved components, personas and backend service options.

% %% ---

\section{Background}
% resource management

\subsection{Resource Management Systems and the SPANK Framework}

Resource management systems in high-performance computing are multi-tenant platforms tasked with allocating compute resources such as nodes, CPU cores, and accelerators according to policies that govern fairness, priority, reservations, and quality of service. These systems manage job submission, queuing, scheduling, execution, monitoring, and completion to ensure efficient and equitable use of cluster infrastructure.

One of the most widely used workload managers in HPC is \emph{Slurm} (Simple Linux Utility for Resource Management), which is an open-source, fault-tolerant, highly scalable system broadly deployed across datacenters, including a majority of the TOP500 supercomputers \cite{slurm_stats, slurm_wikipedia}.

% Slurm’s core architecture comprises:
%
%\begin{itemize}
%  \item a central controller daemon, \texttt{slurmctld}, responsible for scheduling, queue arbitration, and overall cluster state management, with optional automatic failover;
%  \item one or more \texttt{slurmd} daemons executing on compute nodes to launch tasks, monitor execution, and report status back to the controller;
%  \item optional services such as \texttt{slurmdbd} for accounting and \texttt{slurmrestd} for RESTful interactions;
%  \item user-facing commands such as \texttt{sbatch}, \texttt{srun}, \texttt{scancel}, \texttt{squeue}, and \texttt{sinfo} for job submission, control, and monitoring \cite{slurm_overview}.
%\end{itemize}

Slurm supports key functions including exclusive and non-exclusive resource allocation, job execution and monitoring frameworks, and queue-based scheduling that enforces administrative policies \cite{slurm_overview, slurm_wikipedia}. Its modular architecture enables extensibility via plugins, which improves capabilities like accounting, custom resource scheduling, job prioritization, license management, and topology-aware allocation, without altering core source code \cite{slurm_overview}.

One extension mechanism within Slurm is the \emph{SPANK} (Slurm Plug-in Architecture for Node and job Kontrol) framework. SPANK enables developers to implement dynamically loadable, stackable plugins written independently from Slurm’s core which can intercept and modify behavior at specific points in a job's lifecycle: allocation, prologue, task execution, and epilogue \cite{spank_manual}. SPANK plugins are compiled against Slurm’s header file, \texttt{spank.h}, listed in \texttt{plugstack.conf}, and loaded at runtime without requiring modifications to Slurm’s source \cite{spank_manual}. They can register callbacks such as:

\begin{itemize}
  \item \texttt{slurm\_spank\_init}, executed immediately after plugin loading;
  \item \texttt{slurm\_spank\_job\_prolog} and \texttt{slurm\_spank\_task\_init}, triggered before job launch or task execution;
  \item \texttt{slurm\_spank\_job\_epilog}, executed after job completion across different SPANK contexts: \texttt{local}, \texttt{remote}, \texttt{allocator}, \texttt{slurmd}, and \texttt{job\_script} \cite{spank_manual}.
\end{itemize}

This mechanism allows administrators to inject custom behavior such as environment setup, access control, logging, or, in the context of this work, quantum resource management into Slurm's execution workflow.

%\textcolor{red}{[AW: I'd also describe GRES, to foreshadow that]}
Slurm's primary mechanism for managing, node-specific devices like GPUs is the Generic Resource (GRES) scheduling system.
A key distinction between implementing scheduling logic with SPANK or GRES plugins is that GRES plugins are tied to physical resources on a specific node, whereas the SPANK plugin can execute on any node of the system where a job can run.
% Quantum systems

\subsection{Quantum Computing Providers and Architectures}
In the reference implementation we use hardware and software from two different vendors, IBM Quantum and Pasqal,  which we now describe briefly.

\subsubsection{IBM Quantum}

IBM Quantum provides quantum processors built on superconducting qubits. Access is provided through the \textbf{IBM Quantum Platform (IQP)} \cite{ibm_quantum_platform}, a cloud service exposing backends via Qiskit and REST APIs \cite{ibm_quantum_platform}. For on-prem deployments, IBM supports the \textbf{Direct Access API (DA\,API)}, a low-latency interface that enables submission of quantum workloads with parallel lanes for quantum execution.
% note we don't have a reference for DA API... \cite{da_api}.

\subsubsection{Pasqal}

Pasqal builds quantum processors from neutral atoms~\cite{pasqal_architecture}. Pasqal’s native programming library is Pulser~\cite{pulser_lib}.
Access is provided through the \textbf{Pasqal Cloud Service (PCS)}, a cloud service providing access to Pasqal quantum computers and emulators \cite{pasqal_cloud, pasqal_emu}. Pasqal provides an \textbf{on-prem environment} for on-prem QPUs~\cite{pasqal_onprem}, enabling hybrid workflows and concurrent users via a 2-level scheduler.

% Personas

\subsection{Personas}

In the context of quantum-enabled workload management within HPC environments, three primary personas emerge:

\begin{enumerate}
  \item \textbf{HPC users}: These individuals consume compute resources in the cluster and submit hybrid quantum–classical jobs via the scheduler, relying on familiar submission tools (e.g., \texttt{sbatch}). The system abstracts backend complexity from them.
  \item \textbf{HPC administrators}: Responsible for configuring the workload management system, they register quantum resources (e.g., as GRES entries in Slurm), manage access control, and map scheduler-level quantum resources to physical or cloud-based devices.
  \item \textbf{Quantum computer providers}: These entities offer actual QPU execution facilities—either on-site or via cloud APIs (e.g., IBM’s IQP or DA API; Pasqal’s PCS or on-prem solution).
\end{enumerate}

\section{Design}
\label{sec:qrmi_design}
The architectural design for integrating quantum resources into high-performance computing environments should prioritize three key principles: \emph{simplicity}, \emph{loose coupling}, and \emph{long-term adaptability}. An uncomplicated, modular structure reduces integration overhead and encourages maintainability and clarity.

Simplicity enables a clean separation of concerns, such that quantum-specific logic  encapsulated in components like the Quantum Resource Management Interface (QRMI)  remains isolated from workload manager internals and user-facing workflows. This modularity eases component-level testing and future updates.

Loose coupling ensures that modifications to individual system elements whether scheduler internals, plugin logic, or provider interfaces do not necessitate extensive system-wide changes. 

Lastly, the architecture supports \emph{long-term sustainability} through modularity. As quantum hardware and access paradigms diversify, new backends or communication models can be incorporated as plugins or adapters, without altering core scheduler or interface components. This approach follows the ethos of component-based design, enabling evolution without destabilizing existing functionality \cite{modularity, component_based}.

\subsection{Quantum Resource Abstraction Model}

Quantum hardware exhibits variation not only in its physical composition such as individual chips, modules, or integrated QPU clusters but also in how it supports parallel execution, for instance through execution lanes, or partitions. To represent this heterogeneity, we introduce a \emph{quantum resource abstraction}, which encapsulates both the physical quantum device and its concurrency primitives.

Conceptually, a quantum computer is modeled as:
\begin{itemize}
  \item A set of \emph{physical devices} (e.g., individual QPUs, chips, or modules).
  \item A set of \emph{parallelism abstractions} (e.g., execution lanes), enabling the simultaneous submission of multiple quantum tasks; IBM’s execution lanes exemplify this model \cite{qiskit_execution_lanes}.
\end{itemize}

\begin{figure} %[h]
  \centering
  \includegraphics[width=\columnwidth]{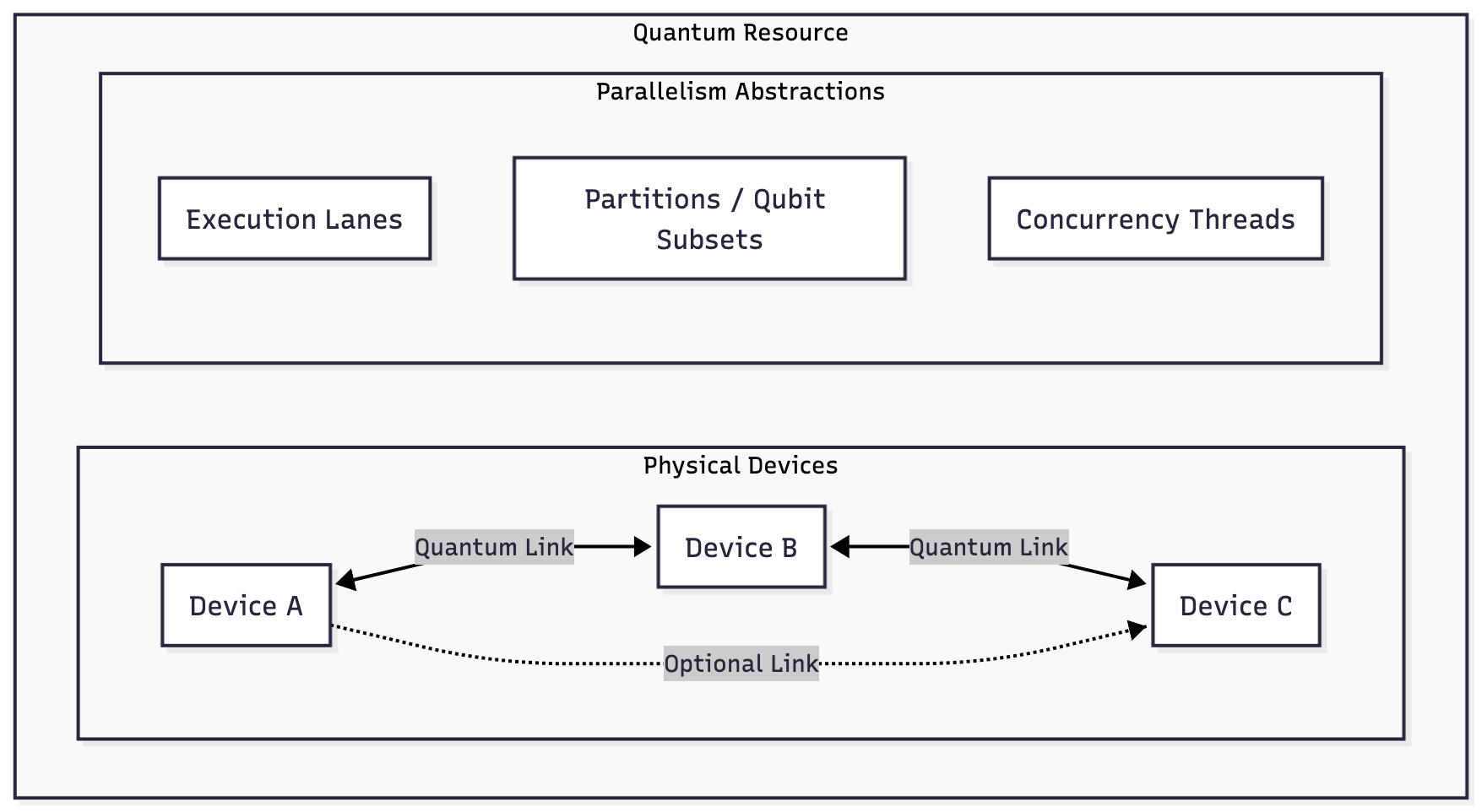}
  \caption{\emph{Quantum resource abstraction model}. A quantum resource can represent a single QPU or multiple QPUs. QPUs can have different properties such as certain connectivity between QPUs, or virtualization/parallelization capabilities like partitions or execution lanes.}
  \label{fig:quantum_resource}
\end{figure}

In practice, a \emph{quantum resource} may correspond to: a single execution lane, a full physical chip, a group of chips, or a partition of a larger device (e.g., qubit subset).

This flexibility allows administrators to tailor resource definitions to each quantum provider’s topology and access semantics. In our implementation, we define a custom resource type commonly `QPU` within the workload management system (e.g., Slurm GRES). This enables mapping scheduler-defined resources to physical systems with precise granularity.

By abstracting away vendor-specific constructs (such as lanes or partitions), the model ensures consistent handling by the scheduler, while preserving deployment flexibility.

\subsection{Quantum Resource Access Model}
\label{quantum_resource_access_model}

\begin{figure}%[h]
  \centering
  \includegraphics[width=\columnwidth]{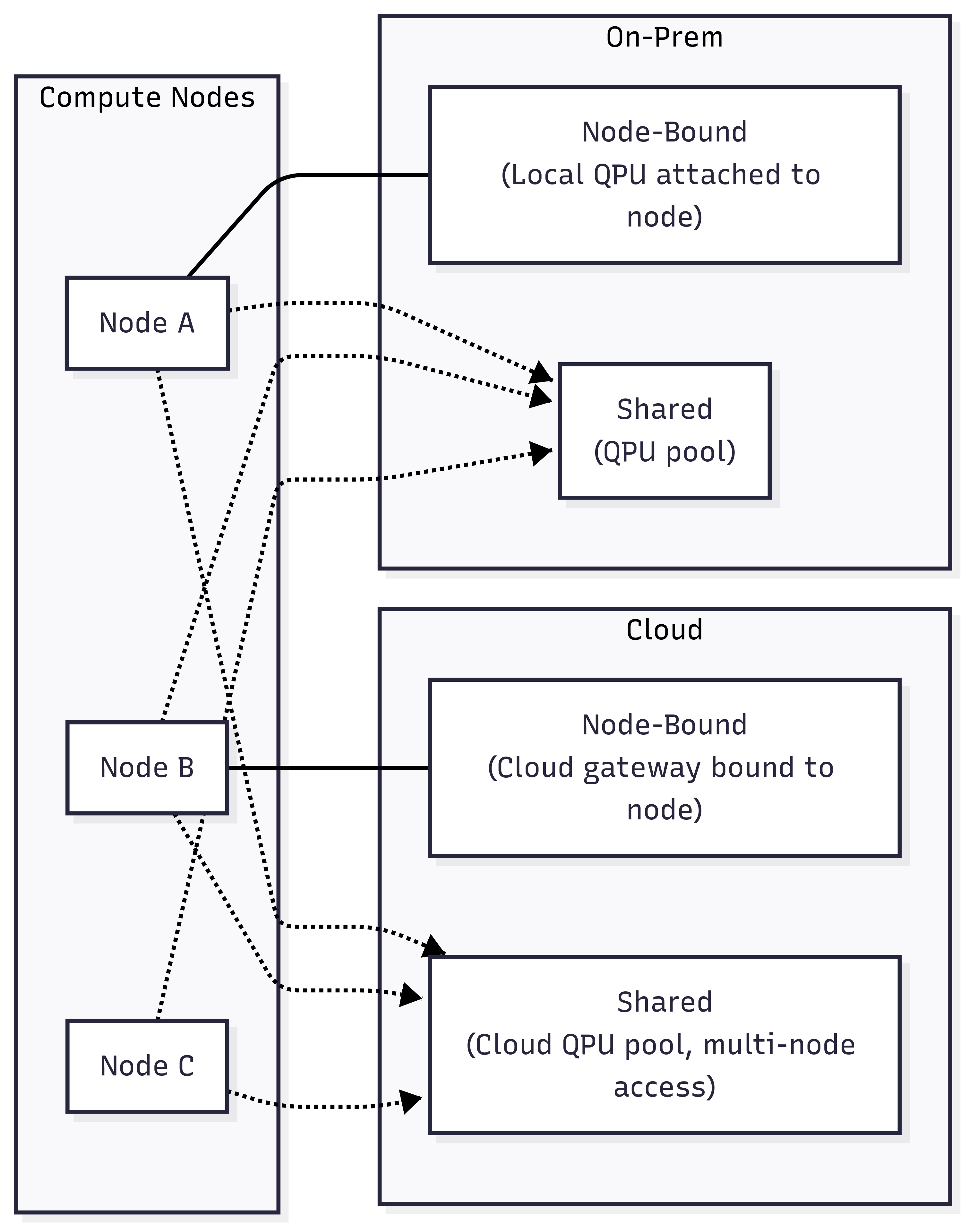}
  \caption{\emph{Quantum resources access model.} Quantum resources can be accessed using direct connections to quantum computer or through cloud APIs. Quantum resources can be node-bound or shared resources accessible by multiple nodes.}
  \label{fig:access_model}
\end{figure}

Quantum-access models vary based on deployment location and infrastructure control. We distinguish between two primary modes:

\paragraph{On-Premises (Direct Access)}  
A quantum device is physically housed within the HPC facility and managed directly by the workload manager. In this configuration, the resource is statically associated with a compute cluster, and acquisition or release of the quantum resource happens in lockstep with job execution. This model minimizes latency with existing security and provisioning frameworks.

\paragraph{Quantum Cloud Bursting}
Quantum devices may also reside in external cloud environments and are allocated dynamically as needed. With this model, the scheduler interacts with cloud-based quantum services (such as IQP or PCS) to obtain quantum resources at job runtime. This aligns with the broader ``cloud bursting'' paradigm offloading excess computational workload to the cloud during peaks - ensuring flexibility and resource elasticity \cite{cloud_bursting}.

These access patterns inform how resources are associated within the cluster:

\begin{itemize}
  \item \textbf{Node-Bound Resources}: The quantum resource is explicitly tied to the compute node where the job runs, typically applicable to on-prem devices co-located with compute hardware.
  \item \textbf{Shared Resources}: The quantum resource is tied to the compute cluster but not to a specific (set of) node(s). In this approach the job can therefore allocate the quantum resource independently of the classical resources. In centers where classical-node-sharing is not possible this enables quantum-resource-sharing nevertheless.
\end{itemize}

\subsection{Quantum Resource Management Interface (QRMI) - Design and Motivation}
\label{qrmi_design}

Integration of quantum resources in HPC environments necessitates a careful architectural strategy. QRMI - the \emph{Quantum Resource Management Interface} - is designed to address the following imperatives:

\begin{itemize}
  \item \textbf{Separation of Concerns}: Workload managers (e.g., Slurm) should remain focused on job and general resource management without embedding quantum-specific behavior, isolating complexity within QRMI.
  \item \textbf{Uniformity Across Vendors}: Users and administrators require consistent behavior regardless of hardware vendor. QRMI abstracts away varying vendor access models.
  \item \textbf{Reusability}: QRMI serves as a pluggable component usable across multiple orchestration systems (e.g., Slurm, Kubernetes), facilitating a “write once, reuse everywhere” strategy. It also
\end{itemize}

QRMI is therefore designed as a thin middleware library, offering a clean and minimal API for scheduler integration composed out of a set of endpoints for resource health check, resource acquisition, task running, hardware topology and monitoring metadata~\cite{qrmi}.
The typical flow with QRMI starts with acquiring (typically locking) a quantum resource first.
Then, jobs can be submitted and execution is monitored.
Eventually, acquired resources are released.
The abstraction of QRMI allows for different implementations, e.g. instead of locking a dedicated backend, the Qiskit Runtime Service implementation for IBM Quantum Platform can create a session with Qiskit Runtime. This allows for continued execution of execution requests, once the session has started to execute.

QRMI offers primitive implementations to the user, which will drive execution of the primitives on the backend. These primitives can vary by type of backend, offering different primitives for e.g. the IBM and Pasqal backends.

% \begin{itemize}
%   \item \texttt{is\_accessible} — check resource availability
%   \item \texttt{acquire} / \texttt{release} — manage resource lifecycle
%   \item \texttt{task\_start}, \texttt{task\_status}, \texttt{task\_result} — control execution and retrieve results
%   \item \texttt{target}, \texttt{metadata} — provide backend-specific hints and resource details
% \end{itemize}

% \textcolor{red}{[IS: add notes on primitives]}

\begin{figure}%[h]
  \centering
  \includegraphics[width=\columnwidth]{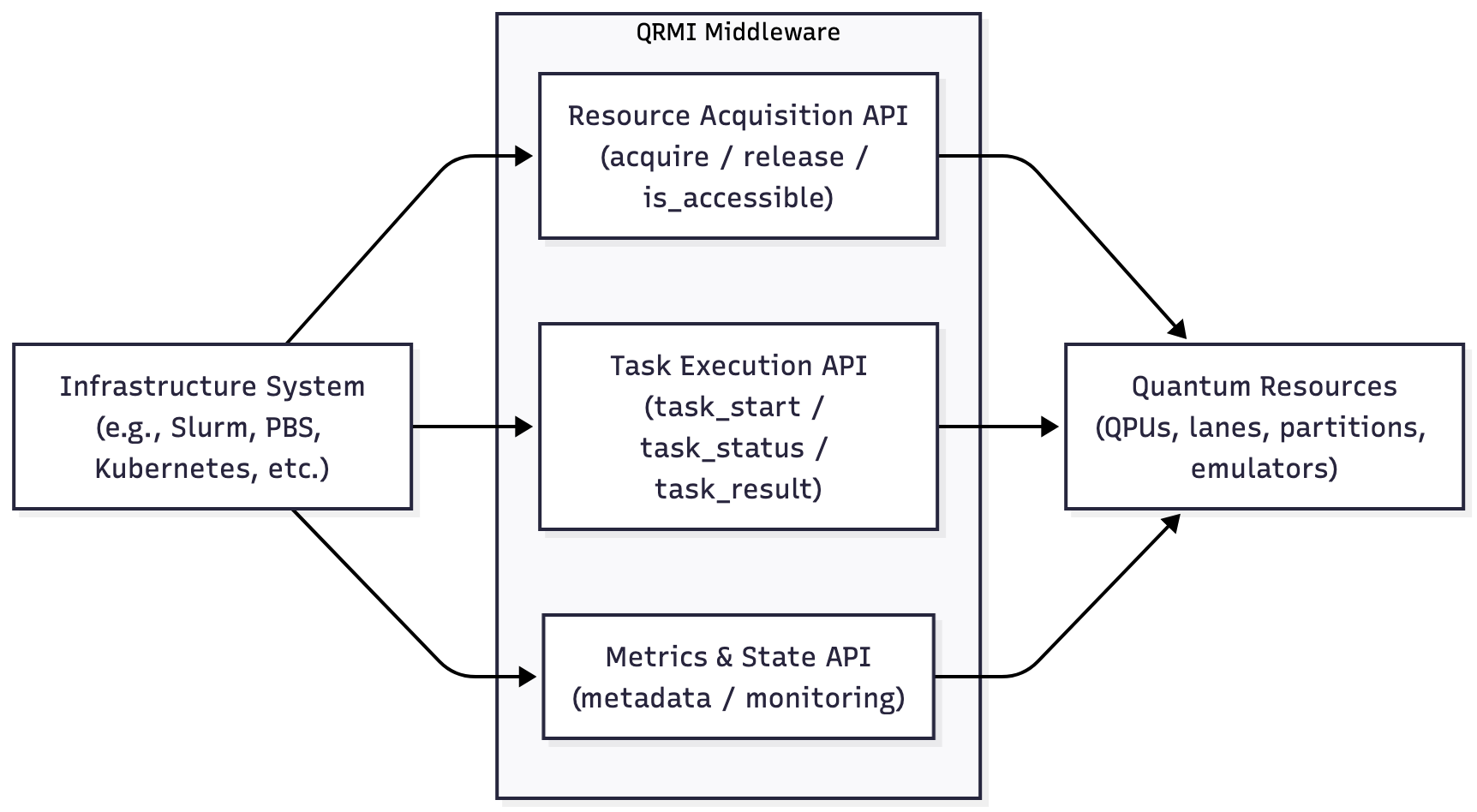}
  \caption{\emph{QRMI overview}. QRMI is a thin middleware layer that contains 3 API groups: resource acquisition, task running and metrics.}
  \label{fig:qrmi_overview}
\end{figure}

This set of abstractions enables scheduler-level code to remain agnostic to vendor details, supporting both on-prem and cloud QPU interfaces through modular backend adapters.

System architecture from a logical perspective follows:
\begin{center}
\texttt{Scheduler / System Software}\\
$\downarrow$ \emph{(calls)}\\
\texttt{QRMI Middleware Library}\\
$\downarrow$ \emph{(dispatches to)}\\
\texttt{Quantum Provider Backend (on-prem or cloud QPU)}
\end{center}

By isolating backend interaction in QRMI, new quantum platforms can be integrated by writing QRMI adapters only, without modifying job schedulers or user interfaces - ensuring long-term maintainability and extensibility.

\subsection{Full System Architecture}

\begin{figure*}%[t]
  \centering
  \includegraphics[width=\textwidth]{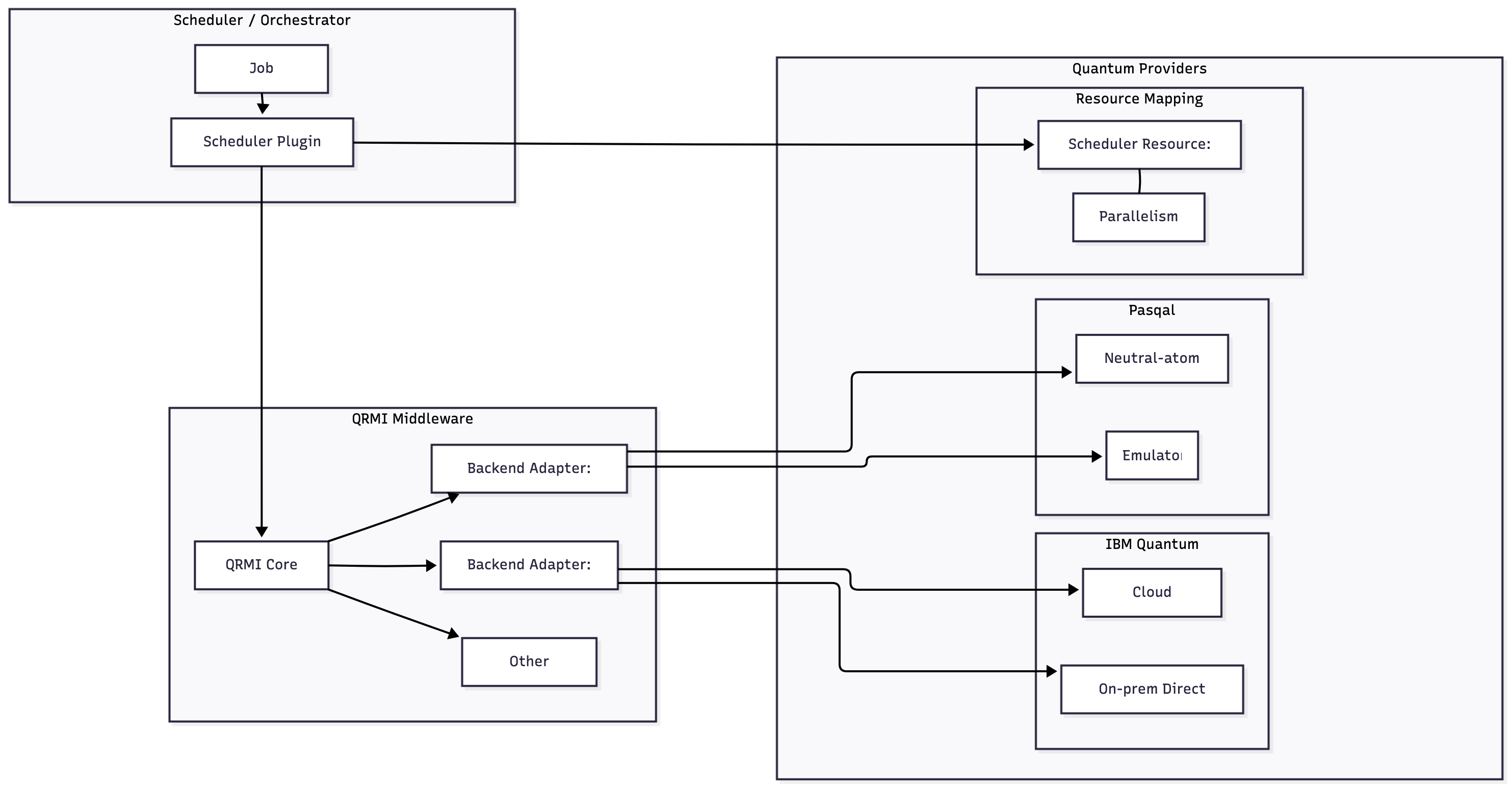}
  \caption{\emph{Workload management system-agnostic integration overview}. Flow for different backends.}
  \label{fig:full_architecture}
\end{figure*}

Figure~\ref{fig:full_architecture} illustrates the high-level architecture for integrating quantum resources into an HPC workload management system - orchestrated by a generalized scheduler.

The main components and flow are as follows:

\begin{enumerate}
  \item \textbf{Custom Resource Definition}: The scheduler (e.g., Slurm) incorporates a new resource type - \texttt{QPU} - which represents quantum hardware resources, including full devices, partitions, or execution lanes.
  \item \textbf{Modular Interaction Flow}: Applications are started as jobs via slurm. A slurm plugin can prepare the information required to access backend services. The application code can then use primitives to trigger execution of circuits. QRMI's primitive implementations allow to appropriately drive execution against the specific backend service. Here, QRMI serves as abstraction for different interfaces and mechanisms of backend services.
  \item \textbf{Lifecycle Integration via QRMI}:
    \begin{itemize}
      \item \texttt{acquire} is called before job execution to reserve the QPU.
      \item Quantum tasks invoke \texttt{task\_start}, \texttt{task\_status}, and \texttt{task\_result} via QRMI.
      \item \texttt{release} is called after job completion to free the resource.
    \end{itemize}
    These calls are embedded via scheduler plugin frameworks (e.g., SPANK).
  \item \textbf{Extensibility}: The scheduler remains unchanged across vendor and infrastructure iterations. QRMI adapters support new resource managers and quantum providers independently.
  \item \textbf{Unified Hybrid Workflows}: Classical and quantum tasks are managed cohesively by the same scheduler, leveraging unified submission, co-scheduling, and resource governance.
\end{enumerate}

\subsection{Limitations and Advanced Capabilities}

While not implemented, a few capabilities are conceivable with this structure

\begin{itemize}
    \item \textbf{Backend sharing:} some backend types like IBM's Direct Access API can parallelize execution preparation, and most cloud-based services also provide a virtualization notion to the submitting entity. It depends on the acquisition mechanism in QRMI whether and how several jobs can be submitted and executed concurrently; the choice of allowing concurrency with other jobs could even be given to users. Additional context is provided in the SPANK plugin documentation~\cite{spank_resource_def}.
    \item \textbf{Resource selection through attributes:} A job submission could choose to have the SPANK plugin select a backend based on criteria (such as qubit count and minimum quality metrics). While this resource selection is not implemented, the SPANK plugin can be extended towards this function, providing greater flexibility and moving away from custom and named backend usage towards a more industrialized usage of quantum computers.
    \item \textbf{Authorization:} the architecture presented in this paper assumes credentials are needed to access backend service, and today's implementation of the SPANK plugin helps to provide this; also, users can override system-wide credentials with their own credentials. However, for backend services that do not offer authentication, the cluster setup needs to consider that access to the service entry point is restricted, and governed through an API gateway component that acts as a gatekeeper.
\end{itemize}

\subsection{Alternative Implementations}

Appendix~\ref{app:alternatives} describes a few additional and alternative implementation possibilities for other scenarios which are not in the scope of this paper.

\section{Implementation}
\subsection{QRMI Implementation}

\begin{figure*}[t]
  \centering
  \includegraphics[width=\textwidth]{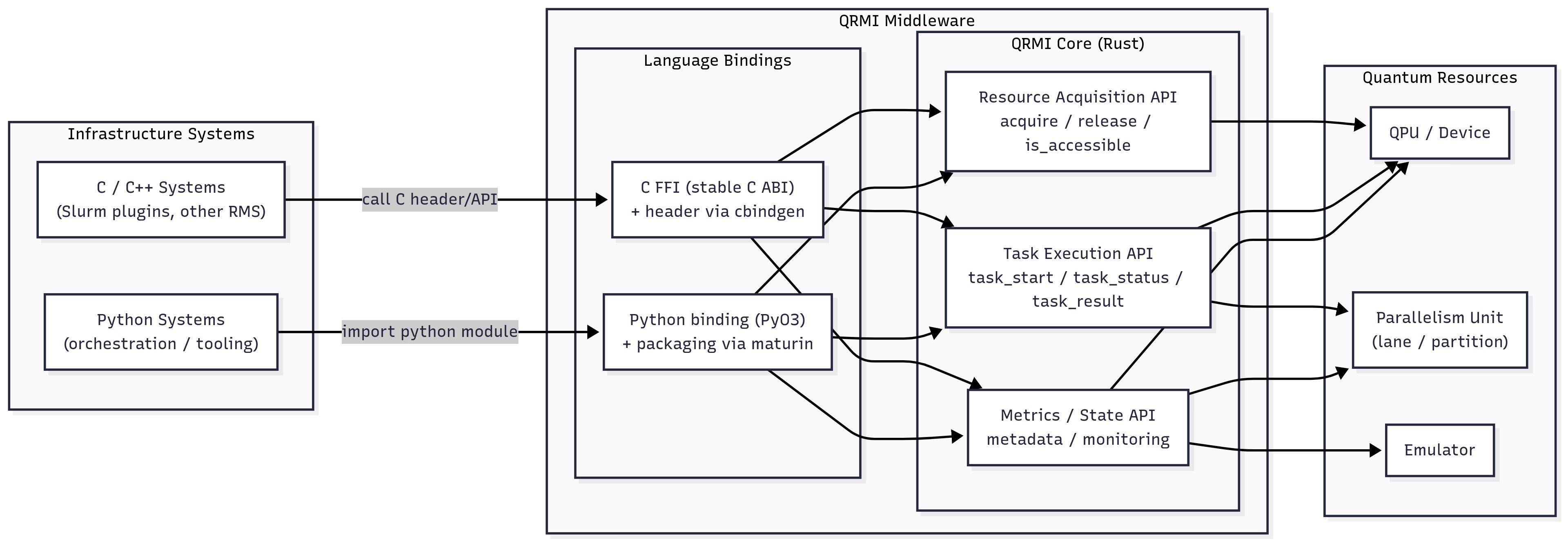}
  \caption{\emph{QRMI implementation}. Implementation of middleware in Rust with C and Python language bindings.}
  \label{fig:qrmi_implementation}
\end{figure*}

The \emph{Quantum Resource Management Interface (QRMI)} is realized as a middleware library that encapsulates quantum resource control, task execution, and monitoring in a vendor-agnostic fashion. The key facets of its implementation are:

\begin{itemize}
  \item \textbf{Core Language and APIs}: QRMI is implemented in \textbf{Rust}, providing strong type safety and performance guarantees, with APIs available for:
    \begin{itemize}
      \item \textbf{C}, for integration with low-level systems.
      \item \textbf{Python}, facilitating scripting and compatibility with tools like Qiskit, Pulser, etc.
    \end{itemize}
  \item \textbf{Monorepo Architecture}: The platform is maintained as a monorepo combining:
    \begin{itemize}
      \item Central QRMI interface module.
      \item Vendor-specific adapters (e.g., IBM Quantum, Pasqal).
      \item Software-stack adapters (e.g. Qiskit, Pulser).
    \end{itemize}
    This ensures synchronized updates, uniform quality controls, and simplified dependency management across providers.
  \item \textbf{Reference Repository}: \url{https://github.com/qiskit-community/qrmi}.
\end{itemize}

QRMI implements the standardized API defined in the design section \ref{sec:qrmi_design} while encapsulating vendor-specific details into isolation modules, thereby achieving modularity, reusability, and maintainability.

\subsection{Reference Integration Architecture: Slurm via SPANK Plugin and GRES}

This subsection outlines how quantum resources are natively integrated into Slurm using the GRES mechanism and SPANK plugins.

\paragraph{Quantum Resource Registration (GRES)}  
Administrators define quantum devices as Slurm Generic Resources—typically labeled "\texttt{qpu}" (or might contain a specific name, e.g. ibm\_torino). Each GRES entry corresponds to a specific quantum backend (e.g., a full chip, execution lane, or partition). This aligns scheduler-level resources with actual hardware capabilities.  

\paragraph{SPANK Hook Integration}  
QRMI actions are embedded into the scheduler via SPANK plugin callbacks at key stages:

\begin{itemize}
  \item \textbf{Job Prologue}: Invokes \texttt{QRMI.acquire()}, establishes middleware, and injects credentials or device identifiers.
  \item \textbf{Task Initialization}: Sets environment variables or context so SDKs (e.g., Qiskit or Pulser) can locate the allocated device.
  \item \textbf{Job Epilogue}: Calls \texttt{QRMI.release()} and performs resource cleanup.
\end{itemize}
This hook-based integration avoids modifying Slurm internals while enabling coordinated quantum resource control through the job lifecycle.

This design ensures unified hybrid job scheduling, embedding quantum resource acquisition and release directly into the standard HPC job submission and lifecycle framework.

With SPANK plugins the availability of the QPU is not accounted for in the scheduling algorithm of Slurm. This can be implemented for example by using the license management feature, or a dedicated queue. Alternatively, QRMI will soon be available with GRES plugin integration, which is taken into account when scheduling. Nevertheless, the SPANK plugin is also of interest once that has been developed as it will not be tied to specific physical nodes. This is especially a constraint in HPC centers where only exclusive access of nodes are present.

\section{Usage}
\subsection{Hybrid Quantum–Classical Workflows}

In an HPC environment augmented with quantum resources, users typically:

\begin{enumerate}
  \item Submit a hybrid Slurm job requesting quantum access, e.g. using \texttt{--gres=qpu:1 --qpu=ibm\_torino}.
  \item Slurm acquires the quantum resource via QRMI before job execution.
  \item Environment variables (e.g., \texttt{QPU\_DEVICE}) are set to guide quantum API calls.
  \item Within the job, user applications perform classical and quantum operations—leveraging frameworks like Qiskit.
  \item Upon completion, QRMI is invoked to release the resource.
\end{enumerate}

\begin{figure*}[t]
  \centering
  \includegraphics[width=\textwidth]{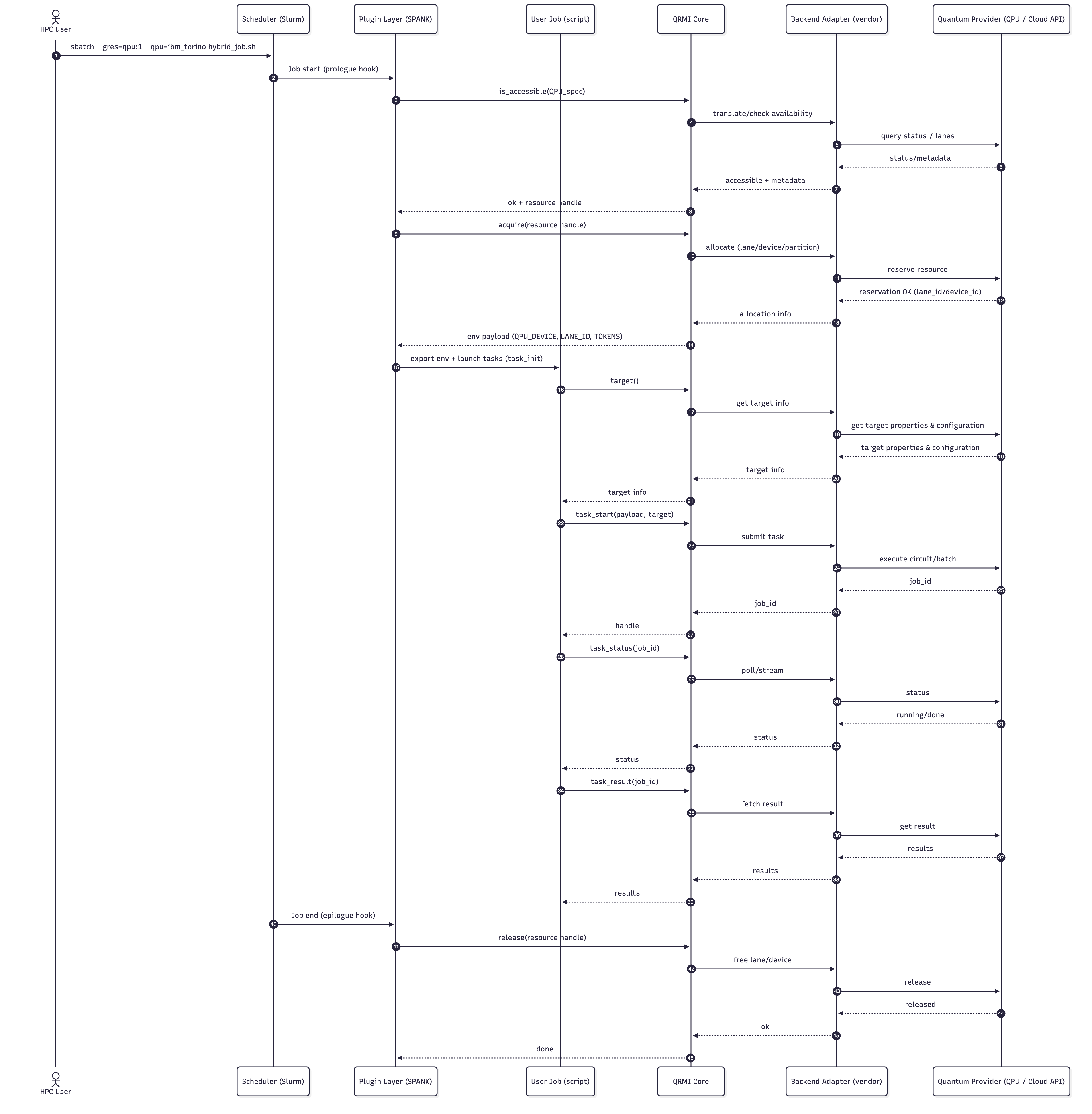}
  \caption{Sequence diagram of hybrid quantum-classical workflow within integrated environment.}
  \label{fig:sequence_workflow}
\end{figure*}

Figure~\ref{fig:sequence_workflow} illustrates sequence diagram of integration and workflow for hybrid quantum-classical environments.

This workflow abstracts away backend allocation logistics, enabling users to focus solely on hybrid computational logic.

To explain the user workflow better, we provide three code examples:

\begin{enumerate}
    \item Slurm submission script for a hybrid job. It provides QRMI with access to two quantum resources
    \item Python script to run on IBM devices using Qiskit and QRMI
    \item Python script to run on Pasqal device using Pulser and QRMI
\end{enumerate}

Below is the example Slurm submission script:

\begin{verbatim}
#!/bin/bash
#SBATCH --job-name=quantum-classical-job
#SBATCH --output=quantum-classical.out
#SBATCH --time=02:00:00
#SBATCH --gres=qpu:2
#SBATCH --ntasks=4
#SBATCH --cpus-per-task=2
#SBATCH --qpu=ibm_fez,FRESNEL

module load python
source venv/bin/activate

python quantum_classical_workflow.py
\end{verbatim}

And a Python script leveraging the quantum resource using Qiskit Sampler:

% somehow lstlisting does not work here?
%\begin{lstlisting}[language=Python]
\begin{verbatim}

import random
import numpy as np
from dotenv import load_dotenv
from qiskit.circuit.library \
  import efficient_su2
from qiskit.transpiler.preset_passmanagers \
  import generate_preset_pass_manager
from qrmi.primitives import QRMIService
from qrmi.primitives.ibm \
  import SamplerV2, get_target

# Create QRMI
load_dotenv()
service = QRMIService()

resources = service.resources()
if len(resources) == 0:
    raise ValueError(
      "No resources available"
    )

# Randomly select a Quantum Resource
qrmi = resources[
  random.randrange(len(resources))
]
print(qrmi.metadata())

# Generate transpiler target 
# from backend configuration & properties
target = get_target(qrmi)

# Create a circuit
circuit = efficient_su2(
  127, entanglement="linear"
)
circuit.measure_all()
# The circuit is parametrized, 
# so we will define the parameter 
# values for execution
param_values = np.random.rand(
  circuit.num_parameters
)

# The circuit and observable 
# need to be transformed 
# to only use instructions
# supported by the QPU 
# (referred to as instruction 
# set architecture (ISA) circuits).
# We'll use the transpiler to do this.
pm = generate_preset_pass_manager(
    optimization_level=1,
    target=target,
)
isa_circuit = pm.run(circuit)

# Initialize QRMI Sampler
options = {
    "default_shots": 10000,
}
sampler = SamplerV2(qrmi, options=options)

# Next, invoke the run() method 
# to generate the output. 
# The circuit and optional
# parameter value sets 
# are input as primitive 
# unified bloc (PUB) tuples.
job = sampler.run([
  (isa_circuit, param_values)
])
print(f">>> Job ID: {job.job_id()}")
print(f">>> Job Status: {job.status()}")
result = job.result()

# Get results for the first PUB
pub_result = result[0]
print(pub_result.data.meas.get_counts())
\end{verbatim}
%\end{lstlisting}

This example shows how the QRMI abstraction uses Pulser for a Pasqal backend:
%This example highlights how scheduler plugins and QRMI abstract away resource management, allowing users to focus on hybrid logic and application-level concerns.

% somehow lstlisting does not work here? verbatim looks nicer
%\begin{lstlisting}[language=Python]
\begin{verbatim}
import json
from dotenv import load_dotenv
from pulser import (
    Pulse,
    Register,
    Sequence,
)
from pulser.backend.remote import (
    JobParams,
)
from qrmi.pulser_backend.backend import (
    PulserQRMIBackend,
    PulserQRMIConnection,
)
from qrmi.pulser_backend.service import (
    QRMIService,
)
from target import get_device

# Create QRMI
load_dotenv()
service = QRMIService()

resources = service.resources()
if len(resources) == 0:
    print(
        "No quantum resource is available."
    )

# Randomly select QR
qrmi = resources[0]

qrmi_conn = PulserQRMIConnection(qrmi)

# Generate Pulser device
device = get_device(qrmi)

reg = Register(
    {
        "q0": (-2.5, -2.5),
        "q1": (2.5, -2.5),
        "q2": (-2.5, 2.5),
        "q3": (2.5, 2.5),
    }
)

seq = Sequence(reg, device)
seq.declare_channel(
    "rydberg", "rydberg_global"
)

pulse1 = Pulse.ConstantPulse(
    100, 2, 2, 0
)

seq.add(pulse1, "rydberg")
seq.measure("ground-rydberg")

backend = PulserQRMIBackend(
    seq, qrmi_conn
)
result = backend.run(
    [
        JobParams(
            runs=1000, variables=[]
        )
    ],
    wait=True,
)
print(
    f"Results: {json.loads(result[0])['counter']}"
)
\end{verbatim}
%\end{lstlisting}

\section{Installations}
This section presents a few examples of slurm installations connecting to Quantum resources, seamlessly integrating quantum compute resources with HPC.

\subsection{RPI CCI AiMOS setup}
Rensselaer's Center for Computational Innovations (CCI) is housed in a 22,000 square-foot facility at the Rensselaer Technology Park. It includes a 4,500 square-foot machine room, offices, and space for industry visitors. The CCI operates heterogeneous supercomputing systems consisting of a massively parallel IBM Data Centric Systems (DCS) supercomputer called {\em AiMOS} and Intel Xeon processor-based clusters. The computational power of {\em AiMOS} is just over 8 petaflops using over 1,500 GPUs on the LINPACK/HPL benchmark and is currently listed as the 166th fastest supercomputer in the world (June 2025, Top500 list) and the 100th most power-efficient supercomputer in the world (June 2025, Green500 list). These systems are supported by over 11 petabytes of disk storage. {\em AiMOS} serves as the test bed for the IBM/NYS AI Hardware Center (see: \url{https://www.research.ibm.com/artificial-intelligence/ai-hardware-center/}), which is a \$2.6 billion program between IBM, New York State, RPI, SUNY Poly and a number of industry partners.

The CCI has dedicated high-speed connections to the main campus and a direct connection to the NYSERNet optical infrastructure and Internet2, providing access to national and international high-speed networks.

\subsubsection*{Specific CCI Resources}
%The CCI Computational facilities include:

\begin{enumerate}
\item {\bf ``AiMOS'' IBM DCS Supercomputer:} 252 compute nodes (14 racks) each with 512 GB of DRAM; 6, NVIDIA Volta V100 GPUs with 32 GB of high-bandwidth memory each; 1.6 TB high-performance NVMe/SSD storage and a dual port 100Gbps EDR InfiniBand card. {\bf Total System: 1512 V100 GPUs and 126TB of memory.}
  
\item {\bf ``NSF MRI'' IBM DCS Supercomputer:} 14 compute nodes (1 rack) each with 512 GB of DRAM; 4 NVIDIA Volta V100 GPUs with 16 GB of high-bandwidth memory each; 1.6 TB high-performance NVMe/SSD storage and a dual port 100Gbps EDR InfiniBand card. {\bf Total System: 56 V100 GPUs and 7 TB of memory.}
  
\item {\bf ``AiMOSx'' HPE GPU/CPU Cluster:} 40 compute nodes each with: 768 GB of DRAM; 8 NVIDIA Volta V100 GPUs with 32 GB of high-bandwidth memory each; 3.2 TB high-performance NVMe/SSD storage and dual port 100Gbps EDR InfiniBand card. {\bf System Total: 320 V100 GPUs and 30 TB of memory.}

\item {\bf Parallel Storage:} 11 petabytes raw disk storage over IBM {\em Spectrum Scale} parallel file system that is able to access data at a peak rate of 80GB per second.
  
\item {\bf Network:} (i) two 648-port non-blocking 100Gbs/EDR Mellanox IB Director switches for AiMOS; and (ii) one, 216-port non-blocking 100Gbps/EDR for other clusters and core networking.

\item {\bf Landing Pads and Front-End Nodes:} The CCI provides 4 ``virtual'' landing pads servers that provide data transfer and two-factor authentication login services. Additionally, AiMOS has two dedicated front-end node systems to support job compilation and job submission via the SLURM job scheduling software.
\end{enumerate}

One of CCI's central strengths is its flexibility in engaging with our industry partners. For example, the CCI can handle corporate confidential data and software.
%\textcolor{red}{IS: TODO: description of CCI and setup}

Connected to this HPC cluster through a dual 10Gb/s link is an IBM Quantum System One, currently with an Eagle processor. Direct Access API is used to access the quantum computer directly without sending circuit inputs and outputs through the Cloud. Slurm and the SPANK plugin are used to run Quantum workload from the HPC cluster.

%\textcolor{red}{IS: TODO: figure AiMOS + system 1 topology, connectivity, throughput}

%\textcolor{red}{IS: TODO: details, lesson learned in text}

\subsection{STFC Hartree Centre cloud quantum-classical setup}

The Hartree Centre deployed the workflow described in this work within a testing environment.
The HPC system used for this project is hosted on Amazon Web Services (AWS) public cloud platform, using their Parallel Cluster service and with the SLURM Scheduler. 
This has made it easy to manage and provision different types of compute nodes and to scale the cluster up and down on-demand. 
For this project, 10 compute nodes have been configured with the “r5.16xlarge” machine type (instances) for running the job workloads. 
These instances are virtual machines that run on Intel Xeon Platinum 8175 processors with a clock speed of 3.1GHz and 32 cores, exposing 64 vCPUs to the virtual machine. 
They each have 512GB of memory and 20 Gigabit networking capabilities; they do not have GPUs or accelerators The cluster is running Red Hat Linux 8.10 operating system, and all nodes have access to a shared AWS EBS file system for storing job scripts and data.
The cluster was then connected with IBM Quantum hardware available via the IBM Cloud service, following the described workflow and procedures outlined in this work.

% ###########
% OLD TEXT
% ###########

% \subsection{Hybrid Quantum-Classical Workflows}

% \textcolor{violet}{IS: TODO: flow diagrams}

% Users can submit jobs that involve both classical and quantum computations.
% The SPANK plugins manage the allocation and execution of quantum tasks, ensuring synchronization with classical components. In most cases, an SDK like Qiskit \footnote{\href{https://github.com/Qiskit/qiskit}{\textcolor[HTML]{840484}{https://github.com/Qiskit/qiskit}}} is used to drive quantum jobs from the classical part of the job.
% This classical job can be integrated into a higher level workflow as is common in HPC projects.

% \subsection{Quantum-Only Jobs}
% For jobs that solely involve quantum computations, the plugins handle the lifecycle of the quantum job, from resource allocation to result retrieval, providing a seamless user experience. These jobs are per se independent of classical processing tied to a slurm job. They can be called in the context of a workflow to ensure consistency in data flow.

% \subsection{RPI CCI AiMOS setup}

% \subsection{STFC cloud quantum-classical setup}

\section{Conclusion}
This work describes a robust approach to integrating quantum computing resources—both on-premise and cloud-based—into existing HPC environments, by treating them as first-class schedulable entities within a workload management system. We have decoupled the quantum resources from the user-SDK in a vendor-agnostic way, implementing SDKs from different vendors.

The reference implementation leverages Slurm along with its SPANK plugin framework, and a thin middleware layer (QRMI), to manage the quantum resource lifecycle—acquire, execute, release—within the native job execution flow. This integration avoids modifications to Slurm’s core components and imposes minimal operational disruption.

The architectural design emphasizes modularity and extensibility. QRMI and the plugin insertion strategy make it feasible to adapt the same pattern to other resource management systems—such as PBS, Kubernetes, etc. — by providing consistent quantum resource handling via middleware and plugins.

In summary, by abstracting quantum-specific complexity behind a minimal interface and integrating it via scheduler hooks, this approach enables coherent, production-ready hybrid quantum–classical workflows within established scientific computing infrastructures.

% \textcolor{violet}{IS: TODO: update}

% The integration of SPANK plugins for quantum resource management within Slurm represents a significant step towards unifying classical and quantum computing workflows. The presented architecture bridges today’s HPC schedulers and emerging quantum accelerators with minimal disruption. By implementing a resource interface in user space and reusing Slurm’s plugin hooks, the solution offers a pragmatic path towards connecting quantum computing and supercomputing. By abstracting the complexities of quantum resource interactions, these plugins enable users to leverage quantum computing capabilities within familiar HPC environments.

%%%==== Added by Stefano
\begin{acknowledgments}
    % Requested by HNCDI
ES, MB, PR and SM work was supported by the Hartree National Centre for Digital Innovation, a UK Government-funded collaboration between STFC and IBM. CDGC and AKK acknowledge support from the project JUNIQ that has received funding from the German Federal Ministry of Research, Technology and Space (BMFTR) and the Ministry of Culture and Science of the State of North Rhine-Westphalia.
\end{acknowledgments}

%%% now appendix also contains alternatives

\appendix
\section{Alternative implementations for HPC workload schedulers}
\label{app:alternatives}

There are a few alternatives to the architecture presented in this paper which are listed for completeness, but not explained in detail:

\begin{itemize}
    \item \textbf{License management:} license management is a cluster-wide way to lock resources, in this case licenses. In a scenario with quantum backends, licenses can serve as a mechanism to lock quantum resources and serialize access to it. Getting a license would correspond to the acquisition step in QRMI. Accordingly the number of licenses can be used to manage how many jobs can submit requests in parallel to the backend.
    \item \textbf{Queues:} additional scheduling queues can be defined, where each queue corresponds to one backend. Any jobs on the queue will execute against the backend. Queue attributes can define how many jobs can run in parallel, i.e. how many applications can submit requests in parallel to the backend; this would correspond to the acquisition step in QRMI. Dependencies can still be established between job of normal and "quantum" queues. This can serve as a simplistic approach e.g. in other schedulers like IBM Spectrum LSF, and does not build on resources as defined e.g. through the SPANK plugin.
    \item \textbf{Native quantum resource implementation:} a new resource type (building on GRES or completely new) can be created in workload schedulers like slurm, creating the desired architecture from scratch. This does not build in the incremental and unintrusive approach of SPANK plugins used in this paper, but instead injects new functionality in the core of Slurm.
\end{itemize}

\section{Quantum Resources for Cloud and Container Management Systems}

Although Slurm (or other workload managers) can be deployed on cloud infrastructure, this approach is not always ideal. Container management systems themselves often serve as resource management frameworks. Therefore, it can be more effective to expose quantum computers as native resources directly within these systems.

\subsection{Quantum Resources for Kubernetes}

In Kubernetes, new types of resources can be introduced as \textit{extended resources} outside of the \texttt{kubernetes.io} domain.%
% cite https://kubernetes.io/docs/concepts/configuration/manage-resources-containers/#extended-resources

To add quantum resource support in Kubernetes, a \textit{Device Plugin} must be implemented. Device plugins allow users to configure clusters with support for resources that require vendor-specific setup, such as GPUs, FPGAs, or, in this case, quantum computers. A device plugin registers itself with the \texttt{kubelet} through the \texttt{Registration} gRPC service. Once registered, users can request quantum resources as part of a Pod specification.

The general workflow of a device plugin is as follows:
\begin{itemize}
  \item \textbf{Initialization:} vendor-specific logic is executed to prepare quantum computers for resource allocation.
  \item \textbf{Service startup:} the plugin starts a gRPC service responsible for handling allocation requests.
  \item \textbf{Registration:} the plugin registers itself with the \texttt{kubelet}.
  \item \textbf{Serving mode:} the plugin continuously monitors device health and reports any state changes to the \texttt{kubelet}. Allocation requests trigger vendor-specific resource acquisition logic.
\end{itemize}

\begin{figure*}[t]
  \centering
  \includegraphics[width=\textwidth]{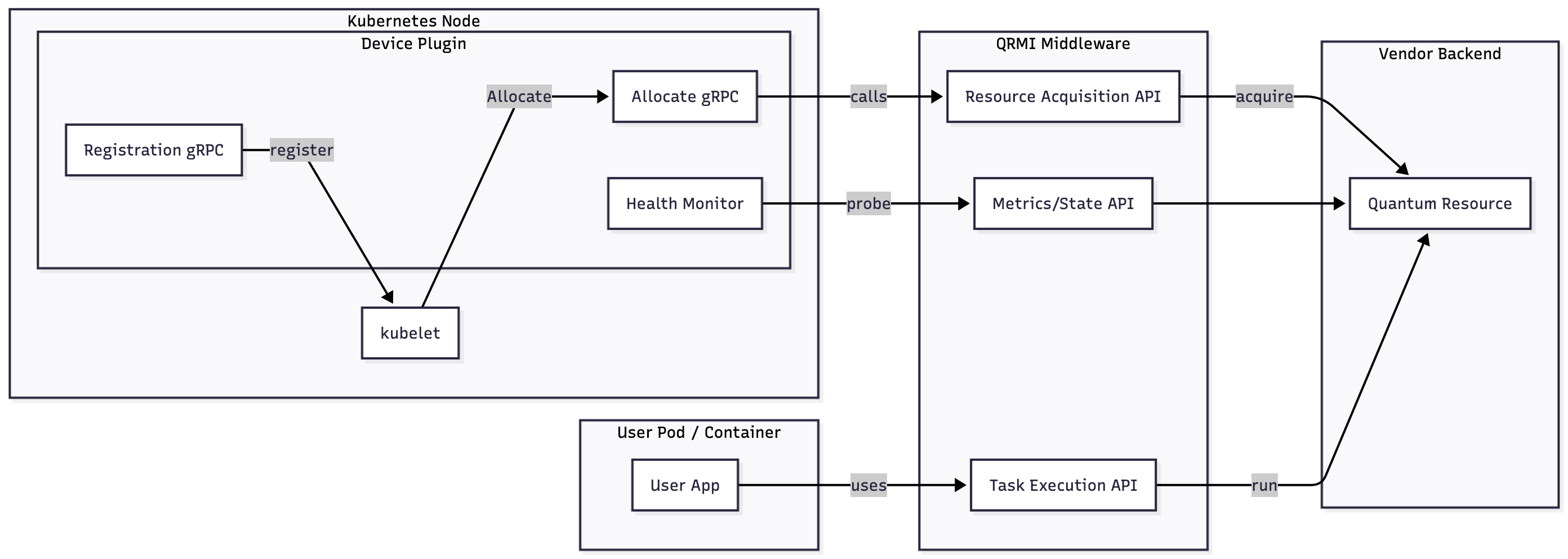}
  \caption{\emph{Kubernetes integration example through Device plugin}. Example diagram showing node level integration to Kubernetes via Device plugin using QRMI library.}
  \label{fig:k8s_implementation}
\end{figure*}

Since QRMI already provides resource allocation logic for multiple vendors, this library can be reused within a Kubernetes device plugin implementation (see Figure \ref{fig:k8s_implementation}). 
% cite https://kubernetes.io/docs/concepts/extend-kubernetes/compute-storage-net/device-plugins/

As discussed in section~\ref{quantum_resource_access_model}, quantum resources can be either node-bound or exposed to the entire cluster as shared resources. For shared resource implementations, the \textit{Dynamic Resource Allocation (DRA)} feature can be utilized. DRA allows resources to be requested and shared among Pods dynamically.%
% cite https://kubernetes.io/docs/concepts/scheduling-eviction/dynamic-resource-allocation/

\newpage
\bibliography{references}

\end{document}